\documentstyle[12pt]{article}
\newcommand{\be}{\begin{equation}}
\newcommand{\ee}{\end{equation}}
\newcommand{\ben}{\begin{eqnarray}}
\newcommand{\een}{\end{eqnarray}}
\begin{document}
\title{Exact Topological Twistons in Crystalline Polyethylene}
\author{E. Ventura and A. M. Simas\\
Departamento de Qu\'\i mica Fundamental, Universidade Federal de Pernambuco\\
50670-901, Recife, Pernambuco, Brazil{\bigskip}\\ and{\bigskip}\\
D. Bazeia$^{\dag}$\\
Departamento de F\'\i sica, Universidade Federal da Para\'\i ba\\
Caixa Postal 5008, 58051-970, Jo\~ao Pessoa, Para\'\i ba, Brazil}
\maketitle

\begin{center}
Abstract
\end{center}

We investigate the presence of topological twistons in crystalline
polyethylene. We describe crystalline polyethylene with a model that
couples the torsional and longitudinal degrees of freedom of the polymeric
chain by means of a system of two real scalar fields. This model supports
topological twistons, which are described by exact and stable topological
solutions that appear when the interaction between torsional and longitudinal
fields is polynomial, containing up to the sixth power in the fields. We
calculate the energy of the topological twiston, and the result is in
very good agreement with the value obtained via molecular simulation.

\vskip 6cm

$^{\dag}${Corresponding author. Fax: +55 83 216 7542;
E-mail: bazeia@fisica.ufpb.br}

\newpage

The existence of twistons in crystalline polyethylene (PE) was postulated
\cite{mbo78} two decades ago, and refers to a twist of 180$^0$
that extends smoothly over several $CH_2$ groups in crystalline PE, in the
plane orthogonal to the chain direction, with the corresponding
$CH_2$ unit length contraction along the polymeric chain. These twiston
configurations appear in crystalline PE as a result of its
large torsional flexibility, and may contribute
to elucidate some of its properties, in particular the dielectric
$\alpha$ relaxation \cite{mrw67,wil72,boy85,wah85,zzo92,zze92}.

There are some interesting models of twistons in crystalline PE
\cite{man80,swo80,spa84,zco93,zco94}. The works \cite{man80,swo80}
are almost simultaneous to the work \cite{ssh79}, which introduces
solitons to describe conductivity in polyacetylene via distortions of the
single-double bond alternations. In the PE chain, however, 
the bounds are always single bounds, which require at least one
bosonic degrees of freedom to describe the torsional flexibility of this
unsaturated polymer. Despite these two decades of investigations,
we believe that the issue of topological twistons playing some role
in explaining properties of the crystalline PE chain is still incomplete,
requiring further investigations both in the theoretical and experimental
grounds. This is the main motivation of the present work, where we follow
an alternate route to topological twistons in PE to bring
new facts to the former theoretical investigations. This new route was
introduced in Ref.~{\cite{bve99}}, and here we complete the investigation,
including the calculation of the energies of the exact topological twistons.

We start our investigations by first reviewing the basic features of the
several distinct mechanical models used to describe twistons in crystalline
PE. The most important ones are described in
Refs.~{\cite{man80,swo80,spa84,zco93,zco94}}. In the pioneer work
\cite{man80} the author considers a system which couples the torsional and
translational degrees of freedom. In Ref.~{\cite{swo80}} the authors
consider a simpler system, describing only the torsional motion along the
crystalline chain, and this is also considered in the subsequent work
\cite{spa84}. It is only more recently \cite{zco93,zco94} that
one includes interactions between radial, torsional and longitudinal degrees
of freedom. In this case one uses cilindrical coordinates to describe a
generic $CH_2$ unit via $(r_n,\theta_n,z_n)$, which correspond to the
three degrees of freedom of the rigid molecular group. A simplification
can be introduced, and concerns freezing the $r_n$'s,
so that the radial motion is neglected. In \cite{zco93} one further
ignores the translational degree of freedom, the $z_n$ coordinates,
to get to a simple model described via the torsional variable that in the
continuum limit can be taken as $\theta_n(t)\to\theta(z,t)$. The model
reproduces the double sine-Gordon model,
according to the assumptions there considered to describe the intermolecular
interaction. The other more recent work \cite{zco94} on twiston in crystalline
PE gives another step toward a more realistic model. This is the first
time the radial, torsional and longitudinal
degrees of freedom are simultaneously considered to model twiston in
crystalline PE. The model is very interesting, although it is hard to
find exact solutions and investigate the corresponding issues of stability.
The problem engenders several intrinsic difficulties, which have inspired us
to search for an alternate model, in the form of two coupled fields belonging
to the class of systems investigated in the recent works
\cite{bsr95,bsa96,brs96,bnt97}.

The basic assumptions introduced in the former
models for twistons in crystalline PE may be described considering
cilindrical coordinates. The Lagrangian
presents the usual form $L=T-U$, where
\ben
\label{ke}
T&=&\frac{1}{2}\,m\,\sum_{n}\,({\dot{r}}_n^2+r_n^2\,{\dot{\theta}}_n^2+
{\dot{z}}_n^2)
\\
\label{pote}
U&=&U_{\rm intra}+U_{\rm inter}
\een
Here $m$ is the mass associated to the molecular group $CH_2$, and
$U_{\rm intra}$ and $U_{\rm inter}$ are potentials used to model
the intramolecular and intermolecular interactions in the crystalline
environment, respectively. The intramolecular potential can be
considered as
\be
\label{intra}
U_{\rm intra}=\frac{1}{2}\,\sum_n\,K_1\,(\theta_{n+1}-\theta_n)^2+
\frac{1}{2}\,\sum_n\,K_2\,(z_{n+1}-z_n)^2+\cdots
\ee
where $K_1$ and $K_2$ are coefficients related to the harmonic
approximation for torsional and longitudinal motions, respectively. The
intramolecular potential may contain derivative coupling between the
torsional and longitudinal motions. In this case we should add to
$U_{intra}$ the contribution \cite{zco93}
\be
\label{div1}
\frac{1}{2}\,\sum_n\,K_3\,(\theta_{n+1}-\theta_n)^2\,(z_{n+1}-z_n)
\ee
However, instead of the above coupling we can consider
derivative coupling between the radial and longitudinal motions. In this
other situation we add to $U_{intra}$ the contribution \cite{zco94}
\be
\label{div2}
\frac{1}{2}\,\sum_n\,K_4\,(r_n-r_0)^2+\frac{1}{2}\sum_n\,K_5\,(r_{n+1}-r_n)
\,(z_{n+1}-z_n)
\ee
The above terms $(\ref{div1})$ and $(\ref{div2})$ are two among several
other possibilities of introducing derivative coupling between the torsional,
longitudinal and radial degrees of freedom. We shall not consider such
possibilities in the present work, although in \cite{bnt97} one shows a route
for taking derivative coupling into account. As we are going to show below,
we consider the standard harmonic approximation for $U_{intra}$ in order to
follow the basic steps of the first works \cite{man80,swo80,spa84} on twistons
in crystalline PE. 

The second potential in $(\ref{pote})$ is $U_{inter}$. It responds for the
intermolecular interactions and is usually given in the form
$U_{\rm inter}=\sum_n\,[U_0(\theta_n)+U_1(\theta_n)\,U_l(z_n)]$. Here
$U_0(\theta_n)$ and $U_1(\theta_n)$ are used to model torsional mobility
and $U_l(z_n)$ to describe the longitudinal motion along the chain. In the
works \cite{man80,swo80,spa84,zco93}, after freezing radial and translational
motion, the above intermolecular potential is described by the
$U_0(\theta_n)$ contributions. We can get to models for the torsional
motion alone, and in the continuum limit they may be described via
the sine-Gordon potential \cite{man80,swo80,spa84}
$A_1\,[1-\cos(2\,\theta)]$, or the polynomial potential \cite{swo80}
$A_2\,\theta^2+B_1\,\theta^4$, or yet the double sine-Gordon
potential \cite{zco93} $A_3\,[1-\cos(2\,\theta)]+B_2\,[1-\cos(4\,\theta)]$.
Here $A_i$ and $B_i$ are real constants, used to parametrize the
corresponding interactions. Evidently, the above potentials lead to different
models for the torsional field, and are introduced to account for the specific
motivations presented in the respective works
\cite{man80,swo80,spa84,zco93}. In the more recent work \cite{zco94}, one
considers coupling between the radial, torsional and longitudinal degrees
of freedom, but the analytical solutions there obtained are found under
assumptions that ultimately decouple the system.

The above models show that the basic idea introduced in
Ref.~{\cite{mbo78}} has survived along the years, although there have been
interesting quantitative contributions to investigate the presence of twistons
in crystalline PE. In particular, in Ref.~{\cite{zco94}} one includes the most
relevant degrees of freedom when one considers the $CH_2$ group in the form of
rigid molecular group along the crystalline chain in crystalline PE. However,
in the model considered in \cite{zco94} we could not fully understand the
reason for not considering harmonic interactions between neighbor radial
coordinates, while taking into account interactions between radial and
longitudinal degrees of freedom in the intramolecular potential.
For this reason, we think we can introduce another mechanical model
for the polymeric chain, where we modify some
assumptions presented in the former works \cite{zco93,zco94}.
The difficulties inherent to the problem of describing topological twistons
in crystalline PE bring motivations for simplifying former assumptions, with
the aim of offering an alternate model that presents exact solutions for
twistons in crystalline PE. Toward this goal, let us use cilindrical
coordinates to describe the molecular groups under the assumption of rigidity.
We start with the kinetic energy $(\ref{ke})$, rewriting it in the form
\be
\label{keo}
T=\frac{1}{2}\,m\,r_0^2\sum_{n}\left({\dot{\phi}}_n^2+
\left(\frac{c}{r_0}\right)^2\,{\dot{\chi}}_n^2+\dot{\rho}_n^2\right)
\ee
Here we have set $\phi_n=\theta_n-[1-(-1)^n](\pi/2)$, $\chi_n=(z_n-nc)/c$
and $\rho_n=(r_n-r_0)/r_0$, where $r_0$ is the equilibrium position
of the radial coordinate and $c$ is the longitudinal
distance between consecutive molecular groups. Now $\phi_n$, $\chi_n$ and
$\rho_n$ are all dimensionless variables, and in the continuum
limit they can be seen as real fields $\phi(z,t)$, $\chi(z,t)$ and
$\rho(z,t)$. Before going to the continuum version of the PE chain,
however, let us reconsider the intramolecular potential given by
Eq.~(\ref{intra}). We use the harmonic approximation to write
\be
\label{intrag}
U_{\rm intra}=\frac{1}{2}\,\sum_n\,k_t\,(\phi_{n+1}-\phi_n)^2+
\frac{1}{2}\,\sum_n\,k_l\,(\chi_{n+1}-\chi_n)^2+
\frac{1}{2}\,\sum_n\,k_r\,(\rho_{n+1}-\rho_n)^2
\ee
where $k_t$, $k_l$ and $k_r$ are spring-like constants, related to the
torsional, longitudinal and radial degrees of freedom, respectively. 

The harmonic interactions present in the
intramolecular term $(\ref{intrag})$ makes the dynamics to appear as the
dynamics of relativistic fields, in the same way it happens with the standard
harmonic chain. We use $(\ref{keo})$ and $(\ref{intrag})$ to write the
following Lagrangian density for the continuum version of the mechanical
model for crystalline PE
\ben
{\cal L}_m&=&\frac{1}{2}\frac{m}{c}r^2_0\,
\left(\frac{\partial\phi}{\partial t}\right)^2-
\frac{1}{2}\,{k_t}{c}\,\left(\frac{\partial\phi}{\partial z}\right)^2+
\frac{1}{2}\frac{m}{c}\,c^2\,\left(\frac{\partial\chi}{\partial t}\right)^2-
\nonumber\\
& &\frac{1}{2}\,{k_l}{c}\,\left(\frac{\partial\chi}{\partial z}\right)^2+
\frac{1}{2}\frac{m}{c}r^2_0\,\left(\frac{\partial\rho}{\partial t}\right)^2-
\frac{1}{2}{k_r}{c}\left(\frac{\partial\rho}{\partial z}\right)^2-
V_{inter}(\phi,\chi,\rho)
\een
The quantity $m/c$ identifies the mass density along the chain, and
$k_tc=\kappa_t,\,h_lc=\kappa_l,\,k_rc=\kappa_r$ are Young parameters
related to the torsional, longitudinal and radial motion, respectively.

The above mechanical model is still incomplete, but it contains the basic
assumption that we are dealing with an harmonic chain, and deviation from
the harmonic behavior is to be included in $V_{inter}$. Although in this case
we can not introduce any other derivative coupling, we still have the freedom
to specify $V_{inter}$ and so introduce nonlinearity via the presence of the
surrounding environment in the crystalline material. This is the model we keep
in mind to introduce the following field theoretic considerations.

We follow the lines of the former mechanical model, which lead us to
introduce a field theoretic model that contains three real scalar fields.
The Lagrangian density describing the fields $\phi=\phi(x,t)$,
$\chi=\chi(x,t)$ and $\rho=\rho(x,t)$ was introduced in Ref.~{\cite{bve99}}.
The model is defined by the potential, which is supposed
to have the form
\be
\label{gpot}
V(\phi,\chi,\rho)=\frac{1}{2}H^2_{\phi}+
\frac{1}{2}H^2_{\chi}+
\frac{1}{2}H^2_{\rho}
\ee
Here $H_{\phi}=\partial H/\partial\phi$ and so forth.
$H=H(\phi,\chi,\rho)$ is a smooth but otherwise arbitrary function
of the fields. This restriction is introduced along the lines of the former
investigations \cite{bsr95,bsa96,brs96}, and leads to interesting
properties, such as the ones explored below.

We focus attention on the crystalline PE chain. In this case it is a
good approximation \cite{man80,swo80,spa84,zco93,zco94} to descard
radial motion in the PE chain. This simplification leads to a system of two
fields, describing torsional and longitudinal motions simultaneously.
However, we first consider the simpler system, described by the torsional
field alone. In this case, in accordance with the Refs.~\cite{zco93,zco94},
investigations on molecular simulation allows introducing the following
torsional potential
\be
\label{p1}
V_1(\phi)=\frac{1}{2}\lambda^2\phi^2(\phi^2-\pi^2)^2
\ee
Fortunately, this potential is generated by the function
$H_1(\phi)=(1/2)\,\lambda\,\phi^2\,(\phi^2/2-\pi^2)$. Also, it has
three degenerate minima, one at $\phi=0$ and the other two at $\phi^2=\pi^2$.

We use the potential $V_1(\phi)$ to get the masses of the elementary
excitations around $\phi=0$ and $\phi=\pm\pi$ in the form
$m_{\phi}(0)=|\lambda|\,\pi^2$ and $m_{\phi}(\pm\pi)=2\,|\lambda|\,\pi^2$.
These results identify an asymmetry in the spectra of excitations of
the torsional motion around the minima $\phi=0$ and $\phi^2=\pi^2$.
This asymmetry appears in consequence of the polynomial potential
$(\ref{p1})$, and is small for small $\lambda$. It is related to the
asymmetry between the well at $\phi=0$, and the well at $\phi^2=\pi^2$.
Since the mass of the field corresponds to the minimum energy
necessary to excitate elementary mesons into the system,
we realize that the value $|\lambda|\,\pi^2$, the difference
$2\,|\lambda|\,\pi^2-|\lambda|\,\pi^2$ may be seen as the energy
for the field $\phi$ go from $\phi=0$ to $\phi=\pm\pi$, that is
the energy to overcome the torsional barrier in this simplified model.

To get to a more realistic model we couple the torsional field to the
longitudinal motion along the chain. We model the presence of interactions
by extending the former function $H_1(\phi)$ to $H_2(\phi,\chi)$ given by
\be
\label{h2}
H_2(\phi,\chi)=\frac{1}{2}\,\lambda\,\phi^2\,(\frac{1}{2}\phi^2-\pi^2)+
\frac{1}{2}\mu\phi^2\chi^2
\ee
This gives the system
\be
\label{1mf}
{\cal L}_2=\frac{1}{2}\left(\frac{\partial\phi}{\partial t}\right)^2-
\frac{1}{2}\left(\frac{\partial\phi}{\partial z}\right)^2+
\frac{1}{2}\left(\frac{\partial\chi}{\partial t}\right)^2-
\frac{1}{2}\left(\frac{\partial\chi}{\partial z}\right)^2
-V_2(\phi,\chi)
\ee
where
\be
\label{p2}
V_2(\phi,\chi)=\frac{1}{2}\lambda^2\phi^2(\phi^2-\pi^2)^2+
\lambda\mu\phi^2(\phi^2-\pi^2)\chi^2+\frac{1}{2}\mu^2\phi^2\chi^4+
\frac{1}{2}\mu^2\phi^4\chi^2
\ee
We are using natural units, as in Ref.~{\cite{bve99}}. The above potential
presents interesting features. For instance,
$V_2(\phi,0)=V_1(\phi)$, which reproduces the torsional model $V_1(\phi)$
when one freezes the longitudinal motion. Also, $V_2(0,\chi)=0$ and
\be
V_2(\pm\pi,\chi)=\frac{1}{2}\mu^2\pi^4\chi^2+\frac{1}{2}\mu^2\pi^2\chi^4
\ee
We can evaluate the quantity $\partial^2 V/\partial\phi\partial\chi$ to see
that it contributes with vanishing values at the minima $(0,0)$ and
$(\pm\pi,0)$. This shows that the spectra of excitations of the
torsional motion around the ground states are unaffected by the presence
of the longitudinal motion. Thus, we can use $V_1(\phi)$ to investigate the
behavior of the torsional motion around the equilibrium configurations.

The masses of the $\phi$ field are now given by
$m_{\phi}(0,0)=|\lambda|\pi^2$ and $m_{\phi}(\pm\pi,0)=2|\lambda|\pi^2$,
around the minima $(0,0)$ and $(\pm\pi,0)$, respectively.
Accordingly, for the $\chi$ field we see that it is massless at
$(\phi=0,\chi=0)$, and at $(\pm\pi,0)$ the mass is
$m_{\chi}(\pm\pi,0)=|\mu|\pi^2$. These results identify
an asymmetry in the spectra of excitations of both
the torsional and longitudinal motion around the minima $(0,0)$ and
$(\pi^2,0)$. This asymmetry appears in consequence of the polynomial
potential $(\ref{p2})$, and is small for small
parameters $\lambda$ and $\mu$. These results allow
introducing the ratio $m_{\chi}/m_{\phi}$ between the masses of the $\phi$
and $\chi$ fields -- see Ref.~{\cite{bve99}}.

The topological solutions connect distinct,
adjacent minima of the potential. The energy corresponding to
the classical configurations can be written in the general form \cite{brs96}
\be
\label{ene}
E_{ij}=|H({\bar{\phi}}_i,{\bar{\chi}}_i,{\bar{\rho}}_i)-
H({\bar{\phi}}_j,{\bar{\chi}}_j,{\bar{\rho}}_j)|
\ee
where $({\bar{\phi}}_i,{\bar{\chi}}_i,{\bar{\rho}}_i)$ and
$({\bar{\phi}}_j,{\bar{\chi}}_j,{\bar{\rho}}_j)$ stand for two vacuum
states, that is, two adjacent points $i$ and $j$ in the field space
$(\phi,\chi,\rho)$ that minimize the potential.

Let us first consider the case of a single field, the $\phi$ field that
describes torsional motion along the polymeric chain. We use former
results to write the equation of motion for static configuration in the form
\be
\frac{d^2\phi}{dz^2}=\lambda^2\phi(\phi^2-\pi^2)(3\phi^2-\pi^2)
\ee
This equation is solved by solutions of the first-order equation
\be
\frac{d\phi}{dz}=\lambda\phi(\phi^2-\pi^2)
\ee
There are topological twistons, given by \cite{bsr95}
\be
\label{s1}
\phi^{(\pm)}_{(t)}(z)=\pm\,\pi\,
\sqrt{(1/2)[1-\tanh(\lambda \pi^2 z)]\,}
\ee
Here we are taking $z=0$ as the center of the soliton, but this is unimportant
because the continuum, infinity chain presents translational invariance.
The sign of $\lambda$ identifies kink and antikink solutions,
connecting the minima $0$ and $\pi$ or $0$ and $-\pi$.
These solutions are stable and can be boosted to their time-dependent
form by just changing $z$ to $\xi=(z-vt)/(1-v^2)^{1/2}$. This model can be
seem as an alternate model to the ones introduced in the former
works \cite{man80,swo80,spa84,zco93}.

The amplitude of the torsional field $\phi$ is $\pi$, which is the angle
the chain rotates to form the twiston. The width of the twiston, $L_{(t)}$,
which is the length along the chain where the angular position of $CH_2$
groups appreciately deviates from the crystalographic positions,
is inversely proportional to the quantity $|\lambda|\pi^2$.
We can also get the energy corresponding to the static twiston.
We use Eq.~(\ref{ene}) to get the value
\be
E_{(t)}=\frac{1}{4}|\lambda|\pi^4
\ee

We now consider the model that describes interactions between the torsional
and longitudinal fields. The equations of motion for static fields
$\phi=\phi(z)$ and $\chi=\chi(z)$ are given by
\ben
\frac{d^2\phi}{dz^2}&=&\lambda^2\phi(\phi^2-\pi^2)(3\phi^2-\pi^2)+
2\lambda\mu\phi(2\phi^2-\pi^2)\chi^2+\mu^2\phi(\chi^2+1)\chi^2
\\
\frac{d^2\chi}{dz^2}&=&2\lambda\mu\phi^2(\phi^2-
\pi^2)\chi+2\mu^2\phi^2\chi^3+\mu^2\phi^2\chi
\een
Although there is no general way of solving these equations, we recognize
that they follow from the potential in Eq.~{(\ref{p2})}, defined via
the function introduced in Eq.~{(\ref{h2})}, and so they are solved by
\ben
\label{foeq21}
\frac{d\phi}{dz}&=&\lambda\phi(\phi^2-\pi^2)+\mu\phi\chi^2
\\
\label{foeq22}
\frac{d\chi}{dz}&=&\mu\phi^2\chi
\een
which are first-order differential equations, easier to investigate.

To find explicit solutions we use the trial orbit method introduced
in Ref.~{\cite{raj79}}. We consider the orbit
\be
\label{o}
\lambda(\phi^2-\pi^2)+\mu\chi^2=\mu(\phi^2-\pi^2)
\ee
We note that this orbit is compatible with the first-order
Eqs.~(\ref{foeq21}) and (\ref{foeq22}). Also,
from Eq.~(\ref{foeq21}) we get 
\be
\label{s21}
\phi_{(t,l)}^{(\pm)}(z)=\pm\,\pi\,\sqrt{(1/2)[1-\tanh(\mu \pi^2z)]\, }
\ee
This result and the orbit (\ref{o}) are now used to obtain, 
\be
\chi_{(t,l)}^{(\pm)}(z)=\pm\,\pi\,\sqrt{\frac{\lambda}{\mu}-1\,}\,
\sqrt{(1/2)[1+\tanh(\mu\pi^2z)]\,}
\label{s22}
\ee
These solutions are valid for $\lambda/\mu>1$ and are similar to the
solutions found in Ref.~{\cite{zco94}} to describe the torsional and
longitudinal degrees of freedom that describe topological twistons in the
crystalline PE chain. 

The amplitude of the twiston is still $\pi$, while the amplitude of the
longitudinal motion is given by $\pi\,[(\lambda/\mu)-1]^{1/2}$.
This result requires that $\lambda/\mu>1$, which is compatible with the
investigation of Ref.~{\cite{bve99}}. In this more sofisticated model
the width $L_{(t,l)}$ of the topological twiston is proportional
to $1/(|\mu|\pi^2)$. It depends inversely on $\mu$. We compare $L_{(t)}$
and $L_{(t,l)}$ to see that $L_{(t,l)}>L_{(t)}$ since $\lambda/\mu>1$ for
the topological twiston of the model of two coupled fields. This result is
new, and shows that the presence of the longitudinal motions contributes
to enlarge the width of the topological twiston. 

Another result follows after calculating the energy of these solutions.
We use Eq.~(\ref{ene}) to get $E_{(t,l)}=E_{(t)}=(1/4)|\lambda|\pi^4$,
which equals the value of the energy of the simpler model, where one
discards the motion of the longitudinal field. This result shows that
although the more general model changes some of the features of
the simpler model, which describes only the twiston field, it does not change
the energy of the twiston. We understand this result as follows: the
first-order equations (\ref{foeq21}) and (\ref{foeq22}) also present
the pair of solutions
\be
\label{s23}
{\bar\phi}_{(t,l)}^{(\pm)}(z)=\pm\,\pi\,\sqrt{(1/2)[1-
\tanh(\lambda \pi^2z)]\, }\;\;\;\;\;\;\;\;\;\;\;\;\; {\bar\chi}_{(t,l)}(z)=0
\ee
This pair of solutions and the former one, given by Eqs.~(\ref{s21})
and (\ref{s22}), are at the same topological sector and present the very
same energy, given in Eq.~(\ref{tle}). However, when one sets $\chi\to0$
in the coupled model, the system changes to the simpler model, and so the
energy of the pair (\ref{s23}) is necessarily equal to the energy of the
twiston in the single field system. This fact explains our results,
and shows that the torsional energy is the main quantity to calculate
the energy of the topological twiston. We use this point of view
to rewrite the energy as $E_{(t,l)}=(1/4)\,(|\lambda|\pi^2)\,\pi^2$.
We have already identified $|\lambda|\pi^2$ and $2|\lambda|\pi^2$
as the masses of the twiston field, which show that when the $\phi$ field
varies from $0$ to $\pm\pi$, that is when a twiston is formed, one changes
from the energy $|\lambda|\pi^2$ to the energy $2|\lambda|\pi^2$,
and this requires the value $|\lambda|\pi^2$. We then identify this
value with the energy for twiston formation along the polymeric chain.
According to Ref.~{\cite{mbo78}}, the energy contribution
of the localized twisted region to the criation of the twist defect is
7.3 Kcal/mol. In fact, in Fig.~[6] and Table II of Ref.~{\cite{mbo78}}
we see that $U_0=9.8-2.5=7.3$ Kcal/mol, which is to be regarded as
the contribution of the localized twisted region to the criation of
the twist defect \cite{boy99}. We then change $|\lambda|\pi^2\to 7.3$
in the energy to obtain
\be
E_{(t,l)}=17.99\, {\rm Kcal/mol}
\ee 
This is the energy of the topological twiston, and is in good
agreement with the energy values of 18.01 Kcal/mol
\cite{mbo78}, 18-19 Kcal/mol \cite{spa84}, and 17.2 Kcal/mol
\cite{zco94}, obtained using different numerical simulations and models.

We conclude this letter recalling that we have investigated
a system of two coupled real scalar fields to model topological twistons
in crystalline PE. This model describes no radial motion, but it couples
the torsional and longitudinal degrees of freedom in a very interesting
way. We have found exact solutions, which engender
several features, and here we offer the following remarks.
The limit $\mu\to\lambda$ transforms the
solutions $(\ref{s21})$ and $(\ref{s22})$ into the solutions $(\ref{s1})$
of the former case, that describes the torsional motion alone.
The solutions of the model of a single field present width proportional
to $1/|\lambda|$, and for the two field model it is proportional to $1/|\mu|$.
The width of the solutions of the two fields is exactly the same,
in agreement with the topological features of the solutions,
and with the orbit (\ref{o}), used to solve the coupled
equations $(\ref{foeq21})$ and $(\ref{foeq22})$.
This result is intuitive, since one expects
that when the torsional motion completes the $180^o$ rotation and returns
to its crystalographic position the longitudinal motion should simultaneously
return to its crystal register. The amplitude of the torsional field is
given by the solution $(\ref{s21})$ and is $\pi$, in agreement with the
model we use for the twiston configuration. The amplitude of the
longitudinal motion is given by the solution $(\ref{s22})$
and is $\pi\,[(\lambda/\mu)-1]^{1/2}$. In the PE chain we have
to set this to unit, to make it compatible with $c$, the full
longitudinal motion. This picture follows in
accordance with the fact that crystalline PE presents
degenerate ground states, obtained from each other by a rotation of 180$^0$
or by a translation of $c$ along the polymer chain.

We have also obtained the energy of the twistons. It is
$E=(1/4)\,|\lambda|\,\pi^4$.
We have used $|\lambda|\,\pi^2$ to identify the mass difference
for the torsional field in the minima $\phi=\pi$ and $\phi=0$. This and
results of Ref.~{\cite{mbo78}} allow getting $|\lambda|\,\pi^2=7.3$ Kcal/mol,
which gives the energy of the topological twiston as $17.99$ Kcal/mol,
in good agreement with values known in the literature.

The results presented in this work completes the former investigation
\cite{bve99}. They show that the approach of using systems of coupled fields
and the corresponding field theoretic analysis to describe
topologically non-trivial excitations in continuum versions of polymeric
chains seems to work correctly. The procedure describes interesting
aspects of the problem, and allows obtaining the energy of the topological
excitation in a direct way. We believe that similar polymeric chains
can also be investigated by similar systems, and this makes us to think
on modelling topological twistons for instance in the
family of systems where one changes some $CH_2$ groups by oxygens
periodically, to make chains with the basic units
$CH_2-O$, $CH_2-CH_2-O$, $CH_2-CH_2-CH_2-O$, etc.
Despite the presence of oxygen the bounds are still sigma bounds,
and the torsional motion seems to be similar to the PE chain. Thus, we may
use a twiston model to explore properties of the family $(CH_2)_n-O$,
in particular in the case of $CH_2-CH_2-O$, the Poly(oxyethylene), POE.
This and other related investigations are presently under consideration.

D.B. and E.V. would like to thank Roman Jackiw and Robert Jaffe for
hospitality at the Center for Theoretical Physics, MIT, where this work has
begun. We would like to thank  R. H. Boyd for the exchange of informations
related to Ref.$\cite{mbo78}$. We also thank  the brazilian agencies
CAPES, CNPq, and PRONEX for partial support.


\begin{thebibliography}{20}
\bibitem{mbo78}M.L. Mansfield and R.H. Boyd, J. Polym. Sci. Phys. Ed.
{\bf 16}, 1227 (1978).
\bibitem{mrw67}N.G. McCrum, B.E. Read and G. Williams, {\it Anelastic and
Dielectric Effects in Polymeric Solids} (Dover, New York, 1991).
\bibitem{wil72}G. Williams, Chem. Rev. {\bf 72}, 55 (1972).
\bibitem{boy85}R.H. Boyd, Polymer {\bf 26}, 323, 1123 (1985).
\bibitem{wah85}K.J. Wahlstrand, J. Chem. Phys. {\bf 82}, 5247, 5259 (1985).
\bibitem{zzo92}G. Zerbi and M.D. Zoppo, J. Chem. Soc. Faraday Trans.
{\bf 88}, 1835 (1992).
\bibitem{zze92}M.D. Zoppo and G. Zerbi, Polymer {\bf 33}, 4667 (1992).
\bibitem{man80}M.L. Mansfield, Chem. Phys. Lett. {\bf 69}, 383 (1980).
\bibitem{swo80}J.L. Skinner and P.G. Wolynes, J. Chem. Phys.
{\bf 73}, 4022 (1980).
\bibitem{spa84}J.L Skinner and Y.H. Park, Macromolecules {\bf 17}, 1735 (1984).
\bibitem{zco93}F. Zhang and M.A. Collins, Chem. Phys. Lett. {\bf 214}, 459
(1993).
\bibitem{zco94}F. Zhang and M.A. Collins, Phys. Rev. E {\bf 49}, 5804 (1994).
\bibitem{ssh79}W.P. Su, J.R. Schrieffer and A.J. Heeger, Phys. Rev.
Lett. {\bf 42}, 1698 (1979).
\bibitem{bve99}D. Bazeia and E. Ventura,
Chem. Phys. Lett. {\bf303}, 341 (1999).
\bibitem{bsr95}D. Bazeia, M.J. dos Santos and R.F. Ribeiro, Phys. Lett.
A {\bf 208}, 84 (1995).
\bibitem{bsa96}D. Bazeia and M.M. Santos, Phys. Lett.
A {\bf 217}, 28 (1996).
\bibitem{brs96}D. Bazeia, R.F. Ribeiro, and M.M. Santos, Phys. Rev.
E {\bf 54}, 2943 (1996).
\bibitem{bnt97}D. Bazeia, J.R.S. Nascimento, and D. Toledo,
Phys. Lett. A {\bf 228}, 357 (1997).
\bibitem{raj79}R. Rajaraman, Phys. Rev. Lett. {\bf42}, 200 (1979).
\bibitem{boy99}R.H. Boyd, private communication.
\end{thebibliography}
\end{document}